\documentclass{sig-alternate}
\usepackage{graphicx}
\usepackage{amsmath}
\usepackage{hyperref}
\usepackage{pgfplots}
\begin{document}
%\conferenceinfo{IIiX 2010,} {August 18--21, 2010, New Brunswick, New Jersey, USA.} 
%\CopyrightYear{2010}
%\crdata{978-1-4503-0247-0/10/08}
%\clubpenalty=10000
%\widowpenalty = 10000

\title{A Subjective Logic Formalisation of the Principle of Polyrepresentation for Information Needs}
\numberofauthors{4} 
\author{
\alignauthor
Christina Lioma\\
       \affaddr{University of Stuttgart}\\
       \affaddr{Stuttgart}\\
       \affaddr{Germany}\\
       \email{liomaca@ims.uni-stuttgart.de}
\alignauthor
Birger Larsen\\
       \affaddr{Royal School of Library and Information Science}\\
       \affaddr{Copenhagen}\\
       \affaddr{Denmark}\\
       \email{blar@db.dk}
\and
\alignauthor 
Hinrich Sch{\"u}tze\\
       \affaddr{University of Stuttgart}\\
       \affaddr{Stuttgart}\\
       \affaddr{Germany}\\
       \email{schuetze@ims.uni-stuttgart.de}
\alignauthor
Peter Ingwersen\\
       \affaddr{Royal School of Library and Information Science}\\
       \affaddr{Copenhagen}\\
       \affaddr{Denmark}\\
       \email{pi@db.dk}
}

\maketitle
\begin{abstract}
Interactive Information Retrieval refers to the branch of Information Retrieval that considers the retrieval process with respect to a wide range of contexts, which may affect the user's information seeking experience. The identification and representation of such contexts has been the object of the principle of Polyrepresentation, a theoretical framework for reasoning about different representations arising from interactive information retrieval in a given context. Although the principle of Polyrepresentation has received attention from many researchers, not much empirical work has been done based on it. One reason may be that it has not yet been formalised mathematically.  

In this paper we propose an up-to-date and flexible mathematical formalisation of the principle of Polyrepresentation for information needs. Specifically, we apply Subjective Logic to model different representations of information needs as beliefs marked by degrees of uncertainty. We combine such beliefs using different logical operators, and we discuss these combinations with respect to different retrieval scenarios and situations. A formal model is introduced and discussed, with illustrative applications to the modelling of information needs.
\end{abstract}

% A category with the (minimum) three required fields
%\category{H.4}{Information Systems Applications}{Miscellaneous}
%A category including the fourth, optional field follows...
%\category{D.2.8}{Software Engineering}{Metrics}[complexity measures, performance measures]



\section{Introduction}
\label{s:Introduction}

Information Retrieval (IR) is primarily concerned with developing models and systems that allow users to find information objects, such as documents, images or videos, which are relevant to their information needs, in an effective and efficient way. Interactive IR (IIR) can be seen as a branch of IR, which considers not only the nature of the retrieval model or retrieval object at hand, but also the different actors of the interaction process, for instance how end-users interact with IR systems and how IR systems should be designed to best support users in the decision processes involved in searching the systems. In a cognitive approach to IIR, the different actors in the interaction process contribute interpretations of their situations and pre-suppositions of the world as well as of the information structures involved. Such interpretations take the form of different representations, for instance, documents, images, music tunes, indexing schemes, retrieval algorithms, as well as user request formulations and work task descriptions representing their information requirements and problem state \cite{IngwersenJ:2005}. In this perspective, such representations, which are manifestations of human cognition, reflection and ideas, can be seen as being contextual to one another, and also as being in a state of interplay with each other over time \cite{LarsenI:2006}.

The principle of Polyrepresentation is a well-known theoretical framework for handling multiple contexts in IR, which contends that the use of cognitively and functionally different representations of information objects in IR may enhance retrieval quality \cite{Ingwersen:1996}. The principle of Polyrepresentation makes deliberate use of a variety of context interpretations by means of the evidence that their representations provide \cite{IngwersenJ:2005}. Specifically, Polyrepresentation encompasses two kinds of representations: 
\begin{itemize}
\item Cognitively different representations, which have been derived from the interpretations by different actors.
\item Functionally different representations, which have been derived from the same actor, such as author generated text structures, image features, diagram captions, and references. 
\end{itemize} 

Central to the principle of Polyrepresentation is the following hypothesis \cite{IngwersenJ:2005}:  
\textit{``The more interpretations of different cognitive and functional nature [...] that point to a set of objects in so-called cognitive overlaps\footnote{Polyrepresentation defines cognitive overlaps as the overlaps of sets of objects created by the divergent cognitive and functional representations.}, and the more intensely they do so, the higher the probability that such objects are relevant (pertinent, useful) to a perceived work task/interest to be solved, the information (need) situation at hand, the topic required, or/and the influencing context of that situation.''} 

Essentially, the principle of Polyrepresentation attempts to make simultaneous combinations of different types of evidence (representative features) that are cognitively contextual to one another, in a structured way. However, even though Polyrepresentation has been proposed more than 15 years ago, relatively few empirically based studies have applied it explicitly for IR purposes (an overview is given in Section~\ref{ss:RelatedWorkPolyrepresentation}). What is currently missing from the principle of Polyrepresentation is an up-to-date flexible mathematical formalisation that will enable the implementation of Polyrepresentation in practice without loss of generality, expressiveness or flexibility of the principle. 

In this work, we propose formalising Polyrepresentation using Subjective Logic, which is a type of probabilistic logic for reasoning in the presence of uncertainty \cite{Josang:2001}. According to the subjective approach to probabilities, the probability of an event is the degree to which someone believes it, as indicated by their willingness to bet or take other actions. This is different to the frequentist approach to probabilities, under which, the probability of an event is the frequency with which it occurs. 

The reason why Subjective Logic is particularly attractive for formalising Polyrepresentation is two-fold. Firstly, Subjective Logic has a very flexible and powerful calculus, which allows us to express representations mathematically with degrees of confidence and uncertainty in the quality of the representation. Practically this means that representations are expressed in a more accurate and specific way, and that any abstractions or estimations we might make when inducing them will be properly reflected formally. Secondly, Subjective Logic has an extended and very powerful set of operators that allow us to combine different representations in fourteen different ways. This flexibility in combining different representations is exactly what has been so far missing from the principle of Polyrepresentation \cite{Larsen:2005}. Different combinations are appropriate for different situations and representations, and we show that Subjective Logic provides a powerful toolset for realising but also reasoning about different combinations. Practically this means that we can select the `most appropriate' combination of representations, where we define `most appropriate' not according to retrieval performance, but according to the features of the representations and of the situation or search scenario at hand.     

The formalisation of the principle of Polyrepresentation with Subjective Logic is realised by making a clear and mathematically tractable analogy between cognitive agents that constitute different representations in Polyrepresentation and opinions that constitute different expressions of belief in Subjective Logic. This work introduces this idea and provides an illustration of its application to the Polyrepresentation of information needs. In addition, two different combinations of representations are presented and discussed, accompanied by illustrative examples. The ideas presented in this paper are not restricted to the specific representations or combinations treated here. The proposed formalism is generic, meaning that any other type or number of representations could be used, and also that different Subjective Logic combination operators could also be applied. 

The main contributions of this work are the following. Firstly, this work bridges two distinct disciplines, namely Information Science and Formal Logic, in order to propose a mathematical formalisation of a well-known principle in cognitive IIR. Secondly, the formalisation of Polyrepresentation using Subjective Logic allows for the first time to model aspects and features of the representations in degrees of uncertainty, in a precise way. This is a departure from the so far Boolean way in which Polyrepresentation has predominantly been practically applied \cite{Larsen:2005}. Finally, the highly expressive calculus of combination operators in Subjective Logic allows to formally select among various different types of combinations, according to the context or situation at hand. This paves the way for selective Polyrepresentation, which is an area that has not received enough attention so far. 

The remainder of this work is organised as follows. Section \ref{s:RelatedWork} overviews past work on two different and until today distinct areas, namely (i) the use of Polyrepresentation in IR, and (ii) the use of formal logic in IR. To our knowledge, there is no past work on formalising Polyrepresentation using logic. Section \ref{s:SubjectiveLogic} introduces the principle of Polyrepresentation on information needs and presents our formalisation of the Polyrepresentation of information needs using Subjective Logic. Section \ref{s:Combinations} discusses the combination of different representations using Subjective Logic operators. Illustrative examples are provided. Section \ref{s:Interaction} discusses the role of interaction in our proposed Subjective Logic framework of Polyrepresentation. Finally, Section \ref{s:Conclusions} summarises our proposal and its contributions, and proposes future extensions of this work.

\section{Related Work}
\label{s:RelatedWork}

\subsection{Polyrepresentation in IR}
\label{ss:RelatedWorkPolyrepresentation}

The principle of Polyrepresentation was initially formulated by Ingwersen (1996), who intended it to be applicable to both exact and best match IR \cite{Ingwersen:1996}. However, in a critical examination of the practical applicability of Polyrepresentation in mainstream IR, Larsen (2005) argued that the principle is inherently Boolean \cite{Larsen:2005}. To remedy that, a \textit{Polyrepresentation continuum} was proposed, as a model for developing the principle towards greater practical applicability \cite{Larsen:2005,LarsenI:2005}. The continuum was defined over two poles, a structured and an unstructured pole respectively. This allowed for any implementation of the Polyrepresentation principle to be discussed in terms of its structure. This Polyrepresentation continuum was further empirically tested by Larsen et al. (2006), and also extended by the addition of a second dimension to it,  which represented query structure and modus \cite{LarsenI:2006}. Based on these extensions, Skov et al. (2008) \cite{SkovL:2008} investigated Polyrepresentation with focus on inter- and intra-document features, in the medical domain. Five functionally and/or cognitively different document representations were identified. The Polyrepresentation hypothesis predicted that combinations of different document representations could lead to improved retrieval performance, compared to single document representations. In this study, combinations of document representations were studied in terms of cognitive overlaps between documents. It was found that the cognitive overlaps generated by combinations of three or four representations of different nature (in accordance with the Polyrepresentation hypothesis) led to more precise retrieval than overlaps generated by (non-Polyrepresentation) combinations of two representations or from single representations. This finding agreed with the predictions made by the principle of Polyrepresentation. This result applied to both structured and unstructured query modes. This study was an investigation of Polyrepresentation on the document-level, in the sense that it studied representations of documents. 

Apart from the document-level, the principle of Polyrepresentation has also been applied on the level of retrieval models, to study different conceptual or algorithmic representations of retrieval models \cite{LarsenI:2009}. In this light, each retrieval model is regarded as a representation of a unique interpretation of IR. The study by Larsen et al. (2009) examined representations consisting of four best-performing retrieval models from TREC-5 \cite{VoorheesB:2005}. These representations were combined according to the principle of Polyrepresentation, which predicted that combinations of very different, but equally good, retrieval models would outperform each constituent as well as their intermediate (non-Polyrepresentation) combinations. An experimental evaluation was carried out against a baseline of eleven different intermediate combinations and also of the four individual retrieval models. Polyrepresentation was overall found to outperform the baselines, albeit with some instability. The central observation drawn from this study was that polyrepresenting retrieval models could be beneficial to retrieval only if the models were quite dissimilar on the conceptual or algorithmical level and equally well-performing, in that order of importance. 

This last condition, namely to combine representations that are equally well-performing, is quite restrictive, in the sense that it prohibits the combination of unequally performing representations that might nevertheless contribute something potentially good to retrieval. More importantly, this condition is highly dependent on how one defines `well-performing'. Different search scenarios, situations and contexts define `good performance' in very different ways. It is exactly this point that we target in this paper, by proposing a flexible Subjective Logic formalism. Specifically, by formalising a representation in degrees of uncertainty, we are allowed to combine different \textbf{and unequal} representations without compromising the outcome of the combination, simply because we can control the contribution of `better' or `worse' representations to their combination. 

In addition to applications of Polyrepresentation on the level of documents and retrieval models, Polyrepresentation has also been used on the query level, and specifically for Interactive Query Expansion (IQE) \cite{DiriyeB:2009}. The main idea behind IQE is to suggest terms to the users during their search in order to enable better retrieval performance. However, the lack of cognitive and functional support during this query refinement process often has counter-effects for the users, who cannot always select terms appropriately, especially when terms are stripped of their context. Diriye et al. (2009) tackled this problem by using Polyrepresentation to improve the presentation of IQE terms to the users. They showed how providing supplementary information on IQE terms can address the ambiguity and uncertainty surrounding IQE, and improve the perceived usefulness of the terms.

Additionally, the principle of Polyrepresentation has recently been applied to create test collections without performing relevance assessments. Efron and Winget (2010) proposed the idea of query aspects based on the principle of Polyrepresentation, and used multiple query aspects to create pseudo-relevance judgments without human intervention. They were able to show that these resulted in a rank ordering of IR systems that correlated highly with rankings based on relevance judgments from human assessors \cite{EfronW:2010}. 

Finally, most recently, Frommholz et al. have been working towards formalising Polyrepresentation within a geometrical IR framework inspired by quantum mechanics~\cite{FrommholzL:2010}. In their model, representations are modelled in Hilbert space, and combined by means of their tensor product. This differs from our work, where representations are modelled as subjective beliefs and combined using logical operators of Subjective Logic. 

\subsection{Formal Logic in IR}
\label{ss:RelatedWorkLogic}

Subjective Logic is a type of formal logic, and more specifically of probabilistic logic, which allows to model degrees of uncertainty about an event (see \cite{Josang:2001} for an introduction, and \cite{Savage:1954} to trace the origins of subjective probabilites). Subjective Logic is not the only formalism to model degrees of uncertainty. Several other mathematical models have been proposed to this end, starting from the Bayesian model of subjective probabilities (see \cite{Fishburn:1986} for a historical survey), as well as generalisations of the Bayesian model (see \cite{Smets:1994} for a survey), the best-known of which is Dempster-Shafer's \textit{Belief Theory}~\cite{Dempster:1968,Shafer:1976} (see \cite{Josang:2001} for an elaborated discussion on the differences between Dempster-Shafer and Subjective Logic, or \cite{LiomaB:2009} for a more succint version).  

One of the earliest applications of formal logic to IR was Van Rijsbergen's \textit{Logical Uncertainty Principle}~\cite{keith:1986}. Various different types of formal logic have since been applied to IR, such as \textit{Modal Logic} \cite{HughesC:1968}, \textit{Situation Theory} \cite{Devlin:1991}, \textit{Plausible Reasoning} \cite{Polya:1954}, and \textit{Terminological Logic} \cite{Baader:1990}, to name some of the main ones. 

Formally, Modal Logic is based on the notion of \textit{possible worlds}, which can be connected to each other via \textit{accessibility relations}. Given a proposition, the evaluation of its truth is realised with respect to a possible world (see \cite{Lalmas:1998} for more). Nie (1992) used Modal Logic to develop a formal IR model, which integrated semantic-based and probabilistic-based approaches of deciding the relevance between a document and a query \cite{Nie:1992}. In Nie's model, documents were seen as worlds, and queries were seen as formulae. A document $d$ was relevant to a query $q$, if $q$ `were true' in $d$ or in a world $d'$ accessible from $d$. This accessibility relation captured documents containing synonymous or near-synonymous terms, or document hyperlinks. 

A variant of the above model for IR was proposed by Chevallet \cite{Chevallet:1992}. This model was formalised using conceptual graphs \cite{Sowa:1984}, which are graphs built out of concepts and their associated semantics. In Chevallet's model, documents and queries were represented by conceptual graphs, and the transformation process was instantiated by operations performed on the graphs. Another variant of the possible worlds formalisation was proposed by Crestani and Van Rijsbergen \cite{CrestaniK:1995}, who used \textit{Logical Imaging}. In their model, terms were seen as worlds, while documents and queries were seen as propositions. A term $t$ `made a document true' if that term belonged to that document. An extension of this model was proposed by Nie et al. (1996), by including user knowledge in the evaluation of the relevance of a document to a query \cite{NieL:1996}. More recently, Zuccon et al. (2008) applied Logical Imaging to IR in terms of Quantum Theory through the use of an analogy between states of a quantum system (i.e. the dynamics of a physical system) and terms in documents \cite{ZucconA:2008}. 

Another type of formal logic used in IR is Situation Theory, which reasons about the concept of information and the manner in which cognitive agents handle and respond to the information picked up from their environment. Situation Theory defines the nature of information flow and the mechanisms that give rise to such a flow by representing information objects as types. Nothing is said about the truth of a type; a type is just the representation of an information object. What makes a type true is the situation (a partially defined world) from which the information represented by that type is extracted. An application of Situation Theory to IR was proposed by Huibers et al. (1996) \cite{HuibersL:1996}. Under this model, a document was seen as a situation $s$ and a query was seen as a type $\phi$. The document was relevant to the query if there existed a flow of information from a situation $s$ to a situation $s'$, such that $s'$ `supported' $\phi$. The nature of the flow depended on the constraints capturing semantic relationships. 

Another type of logic used in IR is \textit{Terminological Logic} \cite{MeghiniS:1993}. In this formalism, documents were represented by individual constants, and classes of documents were represented as concepts. Concepts were characterised by the notion of conceptual containment. Queries were also represented as concepts, hence the retrieval task consisted in finding all documents contained in the concept representing the query.

\textit{Plausible Reasoning} has also been used to develop a logical model for IR by Bruza and van der Weide (1992) \cite{BruzaW:1992}, in an attempt to capture syntactically related information. Documents and queries were represented by index expressions defined upon noun-phrases. The inference process was then based on a series of strict derivation and plausible derivation mechanisms. Each type of derivation came with its own set of rules and axioms. Relevance occured if, given two index expressions $d$ and $q$, representing the document and the query respectively, it could be proven that $d$ `implied' $q$. Somewhat similar to plausible reasoning is \textit{Abductive Reasoning}, which has also been applied to IR using semantic rather that syntactic term relations, by Muller and Kutschekmanesch (1995) \cite{MullerK:1995}. More generally, semantic, syntactic, or other linguistic aspects of information have been modelled by various extensions of formal logic to IR, for instance see Chiaramella and Chevallet (1992) \cite{ChiaramellaC:1992}. 

Particular aspects of formal logic have also been used to address specific aspects or processes in IR, for instance \textit{Belief Revision} has been used to model IR agents~\cite{LoganR:1994}, to estimate the similarity between a document and a query~\cite{LosadaB:2001}, and more recently to model adaptive and context-sensitive IR~\cite{LauB:2008}. Furthermore, \textit{Fuzzy logic} has been used to enhance different stages of image retrieval, such as indexing \cite{SimouA:2008} or relevance feedback \cite{LecceA:2009}, and to model personalised IR \cite{OussalahK:2008}. Among the more recent applications of formal logic to IR, one may note the use of \textit{Description Logic} to model high-precision IR \cite{RadhouaniF:2008} and multimedia IR for educational purposes \cite{LinckelsM:2008}. The Dempster-Shafer theory mentioned earlier remains one of the most widely used logical formalisms in IR. It has been applied to build a complete and holistic IR framework~\cite{Lalmas:1998}, but also to integrate Web evidence into IR~\cite{TsikrikaL:2004}, to integrate evidence of query difficulty into Web IR in the form of semantic scope~\cite{PlachourasO:2005}, as well as to relate dependent indices of IR systems~\cite{ShiN:2008}. Finally, to our knowledge, the only prior use of Subjective Logic to IR has been the recent work of \cite{LiomaB:2009}, who used it to model query difficulty as a subjective belief, formulated on the basis of various types of linguistic evidence.

There exist further applications of formal logic to IR, reviews of which can be found in~\cite{Crestani:2009,Lalmas:1998,keithC:1998}. A more indepth treatment of formal representations for IR can be found in~\cite{keith:2004}.

\section{Modelling Polyrepresentation with Subjective Logic}
\label{s:SubjectiveLogic}

In order to model the principle of Polyrepresentation using Subjective Logic, we create an analogy between representations of information objects (from Polyrepresentation) and subjective beliefs (from Subjective Logic). Under this analogy, polyrepresenting an information object (i.e. combining different representations of an information object) is equivalent to combining subjective beliefs about the truth of that object. Since each representation may have a different degree of uncertainty regarding the enhancement that it may contribute when used for polyrepresentation, we propose to:
\begin{itemize}
\item formalise this uncertainty and consider it in the combination of representations;
\item select different combination operations according to different retrieval contexts and situations.
\end{itemize}

This section explains how we do this, and Section \ref{s:Combinations} discusses the implications of this analogy to IR.  

\subsection{Belief Model with Subjective Logic}
\label{ss:BeliefModel}

The first step towards formalising Polyrepresentation with Subjective Logic consists in defining a belief model. The belief model can be regarded as the setting upon which we will reason about representations as beliefs. 

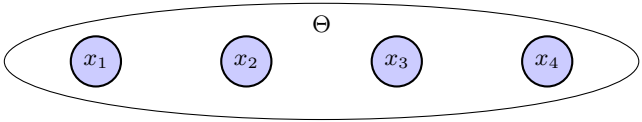
\begin{figure}
\centering
\begin{tikzpicture}
\node at (3,0) [circle,draw=black,fill=blue!20,thick,minimum size=6mm] {$x_1$};
\node at (5,0) [circle,draw=black,fill=blue!20,thick,minimum size=6mm] {$x_2$};
\node at (7,0) [circle,draw=black,fill=blue!20,thick,minimum size=6mm] {$x_3$};
\node at (9,0) [circle,draw=black,fill=blue!20,thick,minimum size=6mm] {$x_4$};
\node at (6,0.5) [] {$\Theta$};	
\draw (6,0) ellipse (120pt and 22pt);
% framed input text
%\draw (12,0) node[text width=3.5cm,fill=blue!10,rounded corners,text justified] {\scriptsize Frame of discernment $\Theta$ with states: $x_1$, $x_2$, $x_3$, $x_4$.};
%\node at (10,0) [] {Example: Frame of Discernment $\Theta$ with four states: $x_1$, $x_2$, $x_3$, $x_4$. The frame is defined over a proposition, which is discerned over the four states.};
\end{tikzpicture}
\caption{\label{fig:example1}
Frame of discernment $\Theta$ with states $x_1$, $x_2$, $x_3$, $x_4$.}
\end{figure}

Belief models define a set of possible situations, for instance a set of possible states of a given system, called \textit{frame of discernment}. This frame is defined over a proposition, i.e. a statement. It is assumed that the system cannot be in more than one elementary state at the same time, or in other words, only one elementary state can be true at any one time. Figure~\ref{fig:example1} illustrates a frame of discernment denoted by $\Theta$ with four elementary states $x_1, x_2, x_3, x_4 \in \Theta$.

Given a frame of discernment over a proposition, one can estimate the probability expectation that this proposition is true. This probability expectation  is computed using evidence, which is said to come from `observers'. An observer can assign to a state a \textit{belief mass}, which represents his belief that this state is true with respect to the proposition. Let $m$ denote belief mass assigned to state $x$ of a frame of discernment $\Theta$, then the following holds:
%\begin{equation}
%m_{\Theta}(x) \ge 0
%\end{equation}
%\begin{equation}
%m_{\Theta}(\emptyset) = 0
%\end{equation}
%\begin{equation}
%\sum_{x \in 2^{\Theta}} m_{\Theta}(x) = 1
%\end{equation} 

\begin{equation}
m_{\Theta}(x) \ge 0 \qquad m_{\Theta}(\emptyset) = 0 \qquad \sum_{x \in 2^{\Theta}} m_{\Theta}(x) = 1
\end{equation} 
\noindent where $2^{\Theta}$ denotes the powerset of $\Theta$.

Subjective Logic considers the belief of an observer about the truth of a proposition as a subjective belief marked by degrees of uncertainty, and it calls it \textit{opinion}. The opinion owner and the proposition are attributes of an opinion. More specifically, an opinion about the truth of state $x$ is defined as the following ordered quadruple:
\begin{equation}
\omega^A_x \equiv (b,d,u,a) 
\end{equation}
\noindent where superscript $A$ is the opinion's owner (i.e the \textbf{observer}), $b$ is the belief mass supporting that the specified proposition is true (i.e. the \textbf{observer's belief}), $d$ is the belief mass supporting that the specified proposition is false (i.e. the \textbf{observer's disbelief})\footnote{The observer's disbelief $d$ corresponds to \textit{doubt} in Shafer (1976)~\cite{Shafer:1976}.}, $u$ is the amount of uncommitted belief mass (i.e. the \textbf{observer's uncertainty}), and $a$ is the apriori probability in the absence of committed belief mass (divided uniformily among the states). It holds that $b+d+u=1$ and $b,d,u,a \in [0,1]$. The uncertainty $d$ of an observer's opinion about the truth of a given state can be interpreted as something that fills the void in the absence of both belief and disbelief. Total uncertainty can be expressed by assigning $b+d=0$. Clearly, an opinion where $b+d=1$ is equivalent to a traditional probability (no uncertainty). The probability expectation\footnote{The probability expectation of an opinion is equivalent to the pignistic probability~\cite{SmetsK:1994}.} of an opinion is: $E=b+au$.

\subsection{Belief Model of Information Needs}
\label{ss:BeliefModelInformationNeeds}

In this work we focus on polyrepresenting information needs. For the purpose of representing an information need as a belief model, we need to take three steps:
\begin{enumerate}
\item Define a frame of discernment over a proposition.
\item Define the states of the frame of discernment.
\item Define the observers who assign opinions about the proposition of the frame.
\end{enumerate}

\subsubsection{Definitions 1 \& 2: Proposition and States}

We define the proposition of our frame of discernment as the abstract information need conceived by the user. This abstract information need is not static, but it tends to alter dynamically \cite{Ingwersen:1996}. In fact, empirical IIR research has shown that the information need perception often changes as a search session progresses (see \cite{Ingwersen:1996} for an overview). If the conception of the information need is seen as a cognitive process, then the changes undergone in this conception can be seen as states in the cognitive process, which we represent as states in our frame of discernment. This allows us to assign belief mass to the abstract information need as a whole ($m(\Theta)$), but also to different states during the cognitive process of the information need ($m_{\Theta}(x)$). This is a flexible and potentially powerful formalism.  For instance, being able to model the temporal cognitive changes in the information need may enable us to use any information extracted from the user over an interactive search session to improve the retrieval results.

\subsubsection{Definition 3: Observers with Opinions}
Having defined the proposition and the states of our frame of discernment, the last step consists in defining the observers who assign opinions to the frame. According to the principle of Polyrepresentation, an abstract information need can have several \textit{concrete} representations, for instance verbalisations by the user, such as keywords. We consider the concrete representations of an abstract  information need as observers of our belief model.  Hence, the opinion of an observer corresponds to how well the concrete representation reflects the abstract information need. More simply, given a user with an abstract information need in his mind, the words he chooses to describe this information need (i.e. his concrete representation of the information need) might not convey perfectly his exact information need as conceived in his mind, because the words he choses might be ambiguous, polysemous, or might convey part but not all of his information need. By representing such a concrete representation as an observer, we can formally represent the departure of the concrete verbalisation of the information need from its abstract conception as the observer's opinion, decomposable into degrees of belief, disbelief and uncertainty. 

For example, a user conceives an abstract information need and verbalises it with the keywords \texttt{health bill US}. For his own reasons (e.g. past use of search engines, exposition to the media, linguistic skills, and so on) the user believes that these three keywords convey his information need. However, these keywords can also be linked to many other different information needs or connotations. It is exactly this `distance' between what the user thinks and how he phrases it that we formally represent as an opinion. The fact that this opinion can be decomposed into elements such as belief, disbelief, or uncertainty, allows us to model the departure of what the user thinks from how he phrases it in a formally tractable way.  

\begin{figure*}
\begin{tikzpicture}
\node at (-6,0.2) [circle,draw=black,fill=blue!20,thick,minimum size=8mm] {$x_1$};
\node at (-4,0.2) [circle,draw=black,fill=blue!20,thick,minimum size=6mm] {$x_2$};
\node at (-5,0.9) [] {\scriptsize \bf abstract information need};	
\draw (-5,0.4) ellipse (30mm and 10mm);
\node at (-11,-2) [fill=gray!3] (A) {\scriptsize \bf Repres. 1};
\node at (-8,-2) [fill=gray!3] (B) {\scriptsize \bf Repres. 2};
\node at (-5,-2) [fill=gray!3] (C) {\scriptsize \bf Repres. 3};
\node at (-2,-2) [fill=gray!3] (D) {\scriptsize \bf Repres. 4};
\node at (1,-2) [fill=gray!3] (E) {\scriptsize \bf Repres. 5};
%\node at (-5,-2.5) [blue,thick] () {\scriptsize \bf OBSERVERS};
\node at (-5,-3.0) [blue,thick] () {\scriptsize \bf (evidence)};

\node at (-8,0) [] (a) {};	
\node at (-7,-0.4) [] (b) {};	
\node at (-5,-0.5) [] (c) {};	
\node at (-3,-0.4) [] (d) {};	
\node at (-2,0) [] (e) {};	

\draw[->,dashed,blue,thick] (A) -- (a);
\draw[->,dashed,blue,thick] (B) -- (b);
\draw[->,dashed,blue,thick] (C) -- (c);
\draw[->,dashed,blue,thick] (D) -- (d);
\draw[->,dashed,blue,thick] (E) -- (e);

\node at (-1,0.9) [blue,thick] () {\scriptsize \bf PROPOSITION};
\node at (1,-0.7) [blue,thick] () {\scriptsize \bf BELIEF MASS};
\node at (1,-1) [blue,thick] () {\scriptsize \bf ASSIGNMENT};

\draw (-5,-3.3) node[text width=17.5cm,fill=blue!10,rounded corners,text justified] {\scriptsize \bf Two states about the proposition of an abstract information need are discerned by the frame. These states represent different stages in the cognitive process of the information need. Five different observers (information need representations) assign belief mass to the frame. Belief mass consists of the observer's belief, disbelief and uncertainty about the truth of the proposition.};
\end{tikzpicture}
\caption{\label{fig:belief-model}A belief model of information needs.}
\end{figure*}

\subsection{Mapping Opinions to Evidence}
\label{ss:EvidenceSpace}

To recapitulate, the previous section defined a belief model based on a frame of discernment, where the user's abstract information need is seen as the proposition of the frame, and the different concrete representations of the information need are seen as observers who have opinions about the truth of the proposition. This belief model is graphically displayed in Figure~\ref{fig:belief-model}. The opinions of these observers are in fact our sources of evidence about the truth of the proposition, or more simply real observations. An observer having an opinion about the truth of a proposition is analogous to the belief, disbelief, or uncertainty of a representation about an abstract information object.

The type of evidence or observations that we use to estimate the truth of the proposition can be seen as either positive (the extent to which the user's concrete verbalisation conveys his abstract information need), or negative (the extent to which the user's concrete verbalisation departs from his abstract information need). Subjective Logic defines a bijective mapping between the opinion and evidence space, as follows. Let $r$ denote positive evidence, and let $s$ denote negative evidence. Then, the correspondence between this evidence and the belief, disbelief, and uncertainty $b,d,u$ is defined as: 
%\begin{equation}
%\label{eq:mapping1}
%b = \frac{r}{r+s+2} 
%\end{equation}
%\begin{equation}
%d=\frac{s}{r+s+2}
%\end{equation}
%\begin{equation}
%\label{eq:mapping3}
%u = \frac{2}{r+s+2}
%\end{equation}
\begin{equation}
\label{eq:mapping}
b = \frac{r}{r+s+2} \qquad d=\frac{s}{r+s+2} \qquad u = \frac{2}{r+s+2}
\end{equation}

\noindent Equation~\ref{eq:mapping} allows one to produce opinions based on statistical evidence. This mapping is derived in a mathematically elegant way (see \cite{Josang:2001} for the full derivation). 

What constitutes positive and negative evidence can be defined in various ways, depending on the situation at hand. For example, let us assume the user who conceives an abstract information need, and who verbalises it using the keywords \texttt{health bill US}. Each of these three keywords can be seen as positive evidence, because they are chosen by the user as adequately representative and salient with respect to his information need. In logic, this is equivalent to saying that the keywords chosen by the user validate the truth of the proposition (i.e. the information need). However, the fact that \texttt{bill} can also convey other senses than the one intended by the user, for instance `invoice', can be seen as negative evidence. In logic, this is equivalent to saying that a keyword conveying a sense other than the one intended by the user does not validate the truth of the proposition (i.e. the information need). In this example, the positive and negative evidence is drawn from lexical-semantic information in a simple way. Evidence can also be drawn from many other types of information, for instance syntactic, statistical, pragmatic or other, as described in \cite{LiomaB:2009}. The choice of evidence used here is illustrative. Any type of contextual information that can potentially be useful may be used as evidence. In addition, the amount of evidence used here is also illustrative. Our belief model allows to represent unlimited additional evidence, simply by introducing more observers who contribute their opinions to the frame of discernment.

\section{Subjective Logic Operations for Combining Representations}
\label{s:Combinations}

Unlike the  example of the previous section, where we had one concrete representation of an abstract information need, namely keywords, in this section we look at information needs for which we have five different representations. These representations are cognitively and/or functionally different, according to the principle of Polyrepresentation. Having different representations, the question is: how best to combine them? We address this question by using various different Subjective Logic operators for combining beliefs. The remainder of this section introduces the different representations of information needs we use, and proposes different operators for combining these representations. 

\subsection{Information Need Representations}
The information needs used here form part of a recent IR test collection called iSearch \cite{LykkeL:2010}. The iSearch test collection is particularly well-suited for testing the polyrepresentation of information needs, since each abstract information need has five different representations:
\begin{enumerate}
\item What the user is looking for.
\item Why the user is looking for this.
\item What is the user's background knowledge of this topic.
\item What should an ideal answer contain to solve the user's problem or task.
\item Which central search terms (keywords) would the user use to express his situation and information need.
\end{enumerate}
\noindent Representation 1 reflects the formulation of the current information need by the user. Representation 2 reflects the user's underlying work task situation or context. Representation 3 reflects the user's current knowledge state. Representation 4 corresponds to the `Narrative' topic field in standard TREC queries. Representation 5 corresponds to the search terms perceived as adequate keywords by the user. The choice of these five representations as adequate reflections of an information need is discussed in \cite{LykkeL:2010} and also in the earlier discussion and experimental evaluation of Kelly \& Fu \cite{KellyF:2007}. Figure~\ref{fig:example2} displays a sample information need from the iSearch test collection with its five representations.
\begin{figure*}
\centering
\begin{tikzpicture}
\draw (-5,0) node[text width=17.6cm,fill=blue!10,rounded corners,text justified] { \footnotesize
\textbf{Representation 1:} \texttt{I am looking for information about manipulation and immobilisation of nano spheres and peptide nano particles.}\\

\textbf{Representation 2:} \texttt{I am starting my master thesis in which I will fabricate self-assembled peptide nano spheres, which needs to be manipulated and immobilized. This is intended done by filling them with metals e.g. gold (Au) or iron (Fe) and use the electrical and magnetic properties to manipulate and immobilise the spheres. This could be by using dielectrophoresis on a chip or micro fluidic device. The nano spheres are intended for biomedical use in which techniques for manipulating biological and biomedical materials are interesting.}\\	

\textbf{Representation 3:} \texttt{The background knowledge is limited since the thesis is starting up this week. But I have been working with sorting of blood cells in micro fluidic devices and flow cytometry.}\\	

\textbf{Representation 4:} \texttt{An ideal answer could be an article showing how to manipulate peptide nano spheres. But in it would in fact might be better if there isn't any articles on the subject since this would mean the research is new.}\\	

\textbf{Representation 5:} \texttt{Manipulation, nano spheres, peptides, immobilisation.}
};
\end{tikzpicture}
\caption{\label{fig:example2}Representations of a sample information need (no. 001) from the iSearch test collection \cite{LykkeL:2010}.}
\end{figure*}

\subsection{Combining Representations}

Having modelled information needs and their representations in a formal way, as shown above, the next step consists in deciding how to combine different representations of the information needs in a flexible and effective way. There exists extended literature on the matter of combining different types of evidence, more formally referred to as fusing beliefs (for instance \cite{Dezert:2002,DuboisP:1988,Josang:2002,JosangD:2010,LefevreC:2002,Murphy:2000,Smets:1990,Yager:1987}, or see \cite{Smarandache:2004} for an overview). Several operators can be used for combining evidence, such as conjunction or disjunction for example. Different operators can produce different results, especially in case of strong conflict between the beliefs to be combined. Furthermore, different operations ought to be used in different situations. We clarify this point with an example, borrowed from \cite{JosangD:2010}.

Let us consider two different situations where we are called to combine beliefs, namely (a) to model the strength of a chain, and (b) to model the strength of a relay swimming team. The correct operator for modelling the strength of the chain is the principle of the weakest link. The correct operator for modelling the strength of the relay swimming team is the average strength of the swimmers. Considering the average strength of the links in the chain to assess the overall strength of the chain might represent an approximation, but it is incorrect, and it could be fatal if life depended on it. Similarly, applying the weakest swimmer principle to assess the overall strength of the relay team might represent an approximation, but it is incorrect, and it would give unreliable predictions. These examples illustrate that situations, which may seem similar at first glance, can be very different when examined more closely, and will therefore require different combination operators. We posit that this applies to different representations of information objects when considering their combination as part of the principle of Polyrepresentation.

Subjective Logic contains fourteen different operators for combining evidence\footnote{A. Josang's draft book in Subjective Logic, available from: http://persons.unik.no/josang/papers/subjective\_logic.pdf}. In this work we illustrate the use of two of these combinations, however the belief model we present allows the use of any of these combination operators. Using Subjective Logic terminology, we will refer to combining evidence as combining opinions, and treat these statements as equivalent.

\subsubsection{Consensus between Independent Opinions}
\label{ss:IndependentConsensus}
The consensus operation of combining opinions  assumes that opinions are independent and that not all the combined opinions have zero uncertainty. Opinions having zero uncertainty would have complete belief or disbelief, and hence would be in complete agreement or disagreement. Attempting to combine solely such opinions using a consensus operator could be seen as meaningless. 

Formally, the consensus operator is defined as follows:  Let $\omega^A \equiv ( b^A, d^A, u^A, a^A)$ and $\omega^B \equiv ( b^B, d^B, u^B, a^B)$ be opinions respectively held by two independent observers $A$ and $B$ about the same proposition. Then, the consensus of opinions of both $A$ and $B$, denoted $\omega^{A,B} = \omega^A \oplus \omega^b$, is defined by:
\begin{equation}
 \label{eq:ind-consensus}
b^{A,B} = \frac{b^A u^B + b^B u^A}{\kappa}
\end{equation}
\begin{equation}
d^{A,B} = \frac{d^A u^B + d^B u^A}{\kappa}
\end{equation}
\begin{equation}
u^{A,B} = \frac{u^A u^B}{\kappa}
\end{equation}
\begin{equation}
a^{A,B} = \frac{a^B u^A + a^A u^B - (a^A + a^B) u^A u^B}{u^A + u^B - 2u^A u^B}
\end{equation}
\noindent where $\kappa = u^A + u^B - u^A u^B$ such that $\kappa \ne 0$, and where $a^{A,B} = (a^A + a^B)/2$ when $u^A, u^B = 1$. The proof is included in ~\cite{Josang:2001}. The effect of the consensus operator is to reduce uncertainty, similarly to Dempster's rule \cite{Dempster:1968} (see \cite{Josang:2001}, Section 5.3, for a discussion on the difference between the two). The consensus operator is graphically displayed in Figure~\ref{fig:consensus}.

Let us contextualise the consensus operator with respect to our information need representations. Consider the information need displayed in Figure~\ref{fig:example2} and let us assume that we draw positive and negative evidence about these information need representations in the simple way presented in Section~\ref{ss:EvidenceSpace} (recall that positive and negative evidence is transformed into the components of an opinion, namely belief, disbelief and uncertainty, using Equation~\ref{eq:mapping}). Let us also assume that these representations are independent, in the sense that they can make sense independently. The combination by consensus would make different sense for representations 1 \& 5 (the description of what the user is looking for \& the keywords), and for representations 3 \& 5 (the user's background \& the keywords), depending on the search scenario at hand:
\begin{itemize}
\item In the case of a standard ad-hoc Web search scenario, where a user is looking for topical information about his conceived information need, the combination operator should consider downweighting the user's background because it might introduce topical drift to the combination. 

\item In the case of a Web search scenario related to professional recruitment or funding, where a user is looking for topical information in combination with his personal skills and background, such as calls for research positions with prerequisites, the combination operator should consider the user's background at least as equally important as the keywords sought. 
\end{itemize}
The above two scenarios illustrate the fact that the consensus combination might not always be the most appropriate way for implementing Polyrepresentation. In addition, a disadvantage of the consensus operator is that it assumes independence of representations. However, this assumption is not always true. On several occasions, representations are not only co-dependent, but also they affect one another. This co-dependence could be modelled in the combination of representations using the recommendation operator of Subjective Logic.

\subsubsection{Recommendation between Dependent Opinions}
\label{ss:Recommendations}

 Let us return to the example information need in Figure~\ref{fig:example2} and let us consider representations 2 \& 4 (why the user is looking for this \& the ideal answer). These representations can be seen as dependent, in the sense that they make better sense when considered together. For example, the fact that the user is starting a Master thesis on a topic clarifies why finding no article on this topic would be ideal. In most occasions, not finding the information sought would be considered a failure, however, for a user looking to conduct novel research, not finding information is a success. This type of user need is highly contextualised. The recommendation operator allows to model this context when combining such representations, in the following way. 
 
 Assume two observers $A$ and $B$, where $A$ has an opinion about $B$, and $B$ has an opinion about a proposition. A recommendation of these two opinions consists of combining $A$'s opinion about $B$ with $B$'s opinion about the proposition\footnote{$B$'s recommendation must be interpreted as what $B$ recommends to $A$, and not necessarily as $B$'s real opinion.} in order for $A$ to get an opinion about the proposition. Formally this is defined as follows: Let $\omega^B \equiv (b^B, d^B, u^B, a^B)$ be $B$'s opinion about a proposition expressed in a recommendation to $A$, and let $\omega^A_B \equiv (b^A_B, d^A_B, u^A_B, a^A_B)$ be $A$'s opinion about $B$'s recommendation. Then, the combination by recommendation $\omega^{AB} = \omega^A_B \otimes \omega^B$ is $A$'s opinion about the proposition as a result of the recommendation from $B$, defined as:
\begin{equation}
 \label{eq:recommendation}
b^{AB} = b^A_B b^B
\end{equation}
\begin{equation}
d^{AB} = b^A_B d^B
\end{equation}
\begin{equation}
u^{AB} = d^A_B + u^A_B + b^A_B u^B
\end{equation}
\begin{equation}
a^{AB} = a^B
\end{equation}

The recommendation operator is graphically displayed in Figure~\ref{fig:recommendation}.

The recommendation opinion is powerful in the sense that a given recommendation can be allowed to bias significantly the combination. In the example used above, biasing the combination in favour of the user's reasons for searching (representation 2) would indicate that the option of retrieving no information would be better than the option of presenting the user with peripheral but not relevant information. Since this is the opposite of what most search engines assume and implement, such contextual knowledge could be valuable. Finally, note the inadequacy of the consensus operator in this scenario. 
\begin{figure}
	\begin{tikzpicture}
		%\node at (-7,5.0) [circle,draw=blue,fill=white,thick] {};	
		\node at (-7,4.6) [black,thick] (state1) {\scriptsize \bf proposition};
		\node at (-10.0,1.0) [] (A) {\scriptsize \bf observer A};
		\node at (-4.0,1.0) [] (B) {\scriptsize \bf observer B};		
		\draw[-> ,dashed,black,thick] (A) -- (state1);
		\draw[-> ,dashed,black,thick] (B) -- (state1);
		\node at (-7.0,1.0) [blue] (C) {\scriptsize \bf A,B};		
		\node at (-7.0,0.6) [black,thick] () {\bf $\omega^A \oplus \omega^B$};
		\draw[-> ,dashed,blue,thick] (A) -- (C);
		\draw[-> ,dashed,blue,thick] (B) -- (C);			
		\draw[-> ,blue,thick] (C) -- (state1);
		\draw (-4,4.0) node[text width=2.5cm,fill=blue!10,rounded corners,text justified] (arrow-begin) {\scriptsize \bf combined opinion};
		\node at (-7.1,4.0) [] (arrow-end) {};
		\draw[- ,black,thick] (arrow-begin) -- (arrow-end);
		\end{tikzpicture}
		\caption{\label{fig:consensus}Combining the opinions of two independent observers using consensus.}
\end{figure}
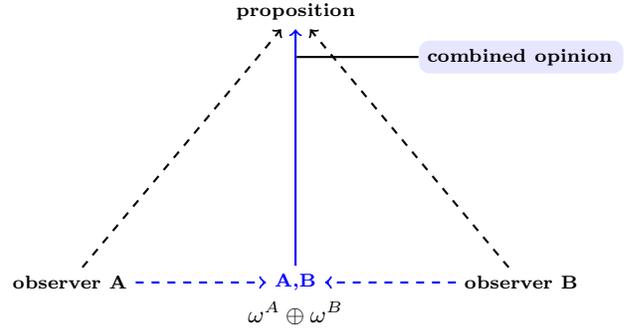
\begin{figure}
	\begin{tikzpicture}
		\node at (-7,4.6) [black,thick] (state1) {\scriptsize \bf proposition};
		\node at (-10.0,1.0) [] (A) {\scriptsize \bf observer A};
		\node at (-4.0,1.0) [] (B) {\scriptsize \bf observer B};		
		\draw[-> ,dashed,black,thick] (A) -- (B);
		\draw[-> ,dashed,black,thick] (B) -- (state1);
	
		\draw[-> ,blue,thick] (A) -- (state1);
		\node at (-10.0,3.0) [black,thick] () {\bf $\omega^A \otimes \omega^B$};
		\draw (-4.0,4.0) node[text width=2.5cm,fill=blue!10,rounded corners,text justified] (arrow-begin) {\scriptsize \bf combined opinion};
		\node at (-7.6,4.0) [] (arrow-end) {};
		\draw[- ,black,thick] (arrow-begin) -- (arrow-end);
		\end{tikzpicture}
		\caption{\label{fig:recommendation}Combining the opinions of two dependent observers using recommendation.}
\end{figure}
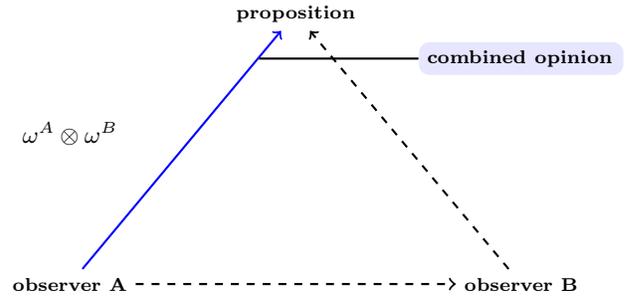
%\begin{table}
%\centering
%\caption{\label{tab:p10}Precision@10 for query no. 42. Retrieval with Language Model (Kullback-Leibler with Jeliner-Mercer smoothing, $\lambda$ tuned for P@10) using Lemur.}
%\begin{tabular}{|l|c|} 
%\hline
%Representation	&P@10\\ 
%\hline
%\hline
%Keywords 	&0.1\\
%Work Task	&0.4\\
%Background	&0.2\\
%\hline
%Keywords+Work Task	&0.4\\
%Keywords+Background	&0.2\\
%Work Task+Background	&0.5\\
%\hline
%\end{tabular}
%\end{table}

\section{The Role of Interaction in our Polyrepresentation Framework}
\label{s:Interaction}
So far we have presented a Subjective Logic formalisation of the principle of Polyrepresentation. A very important aspect of this framework is the process of IR interaction, because this process provides a range of elements that constitute additional representations of the retrieval situation, or more simply the Polyrepresentation input. 

In the user scenarios presented so far, such elements are, for instance, various kinds of implicit and explicit relevance feedback information from the user to the system, based on input data from the system. Additionally, in other scenarios, such elements could be explicit user perceptions or opinions given at search time or over time, such as ratings, recommendations, tags, citations, etc. From the IR system side, such elements may allow to build a polyrepresentative user model - for the purpose of social or individual IR personalisation. From the user side, the output of the IR system through the interaction iterations may provide users with a polyrepresentation of the `system' (e.g., the IR engine, its algorithms, errors, bias, features of the retrieval collection, and so on). An example of this is the classic form of user query modification, through suggested term lists, snippets, concept structures, ontology representations, etc. All of these can be seen as different representations made from differing perspectives and interpretations of the same body of information objects or algorithms. 

In the present work the focus is on establishing a formalism that allows a mathematically tractable and flexible combination of a wide range of representations that occur in IIR scenarios. How to extract these representations and how to make use of them in an interactive setting is the subject of future work.

\section{Conclusions}
\label{s:Conclusions}

This work presented a logical formalisation of the principle of Polyrepresentation for information needs in the context of Interactive Information Retrieval (IIR). IIR is a branch of IR that considers, not only the nature of the retrieval model or retrieval object, but also the different actors of the whole information seeking \& interaction process. The logical formalism introduced in this work was based on Subjective Logic, a formal calculus for representing and operating on probabilities in the presence of uncertainty. We introduced a belief model for reasoning about different representations of information needs in terms of beliefs marked by degrees of uncertainty. We discussed the combination of different representations with respect to different scenarios and contexts, and we presented different formal operators for combining representations in these different scenarios. 

The flexibility of this formalisation paves the way for further research into more structured and selective approaches to Polyrepresentation, not only in the modeling of representations, but also in their combinations. Furthermore, we intend to work towards deriving general principles for suggesting specific combinations of representations. Such principles could underpin our proposed selective approach to Polyrepresentation, and would enable grounding this approach to empirical data and hence rendering it operational. 

\section{Acknowledgments}
Work partly funded by the Research Council of the Ministry of Culture, Denmark (grant number: 2008-001573 - project entitled \textit{Relevance in Context}.)

\bibliographystyle{abbrv}
\bibliography{PolyrepSubjLog}  
\end{document}